\documentclass[12pt]{article}
\usepackage[numbers]{natbib}
\usepackage{amsmath, amssymb, amsthm, geometry}
\title{Curvature-Induced Saturation in Catalytic Reaction Networks: A Differential Geometrical Framework for Modeling Chemical Complexity}
\author{Anderson M. Rodriguez}
\date{April 2025}

\begin{document}

\maketitle

\begin{abstract}
The evolution of chemical reaction networks is often analyzed through kinetic models and energy landscapes, but these approaches fail to capture the deeper structural constraints governing complexity growth. In chemical reaction networks, emergent constraints dictate the organization of reaction pathways, limiting combinatorial expansion and determining stability conditions. This paper introduces a novel approach to modeling chemical reaction networks by incorporating differential geometry into the classical framework of reaction kinetics. By utilizing the Riemannian metric, Christoffel symbols, and a system-specific entropy-like term, we provide a new method for understanding the evolution of complex reaction systems. The approach captures the interdependence between species, the curvature of the reaction network's configuration space, and the tendency of the system to evolve toward more probable states. The interaction topology constrains the accessible reaction trajectories and the introduced differential geometrical approach allows analysis of curvature constraints which help us to understand pathway saturation and transition dynamics. Rather than treating reaction space as an unconstrained combinatorial landscape, we frame it as a structured manifold with higher order curvature describing a geodesic for system evolution under intrinsic constraints. This geometrical perspective offers a unique insight into pathway saturation, self-interruption, and emergent behavior in reaction networks, and provides a scalable framework for modeling large biochemical or catalytic systems.
\end{abstract}

\section{Introduction}
Chemical reaction networks are traditionally analyzed using kinetic models, which focus on reaction rates, species concentrations, and the dynamics of system evolution over time. Kinetic models, based on mass action or more complex formulations, provide valuable insights into the behavior of reaction networks under various conditions. However, traditional kinetic models often fail to capture certain complex behaviors, such as pathway saturation, nonlinear evolution, and system constraints as the number of reactions grows. Specifically, they often assume an unconstrained combinatorial expansion of possible reaction pathways, which does not reflect the inherent limitations present in real systems.

In this paper, we introduce a novel approach to understanding chemical reaction networks by incorporating differential geometry. Differential geometry is the study of smooth manifolds which are shapes and spaces that are differentiable at all points. In this paper we seek to utilize analytical tools from the discipline in regard to analysis of the abstract higher order configuration space of some arbitrary chemical reaction--the theory being that the curvature of the system is defined in relation to the first order solution by acting as a global configuration constraint. i.e., the system is not infinitely combinatorial but is in fact constrained by the bounds of possible interactions given by the structure of the reaction space.  

This approach aims to capture nonlinear evolution and saturation dynamics that are neglected or difficult to analyze through traditional kinetic methods. The primary tool we use shall be a Riemannian manifold, a geometric framework that allows us to define concepts such as distance, volume, angles, etc. in the context of the system. The Riemannian metric on this manifold provides an inner product for each tangent space, enabling analysis of the geometric constraints on the system's evolution.

\section{Curvature as Structuring Constraint in Catalytic Networks}
Consider a system of \( N \) chemical species participating in a catalytic reaction network, where each reaction obeys conservation laws and intermediate stability conditions. We define the system state as a vector \( x^i \in \mathcal{C} \), where \( \mathcal{C} \) represents the configuration space of reactant concentrations and kinetic parameters.

As we shall see, the curvature defined in the configuration space represents how the reaction rate changes as a function of concentration, effectively governing the pathway saturation we observe in real chemical systems, i.e., curvature in the reaction space limits the expansion of reaction pathways, forcing the system to saturate once critical complexity is reached.

Just as in chemical equilibrium, curvature limits the number of feasible reaction pathways, preventing an infinite combinatorial explosion of possible reactions.

\subsection{Constraint Manifold and Curvature-Induced Pathway Saturation}
The first step in our framework is defining a Riemannian metric \( g_{ij} \), which governs the geometry of the system's reaction space. The term \( g_{ij} \) is the inner product between tangent vectors \( i \) and \( j \). The term \( g_{ij} \) describes how species' concentrations influence each other, similar to how a change in one reactant's concentration can shift the rates of other reactions.

We impose a Riemannian metric on \(\mathcal{C} \) of the form:
\begin{equation}
    g_{ij} = \alpha \delta_{ij} + \beta f_{ij}(x),
\end{equation}

Where:
\begin{itemize}
  \item \( \alpha \) is a constant that scales the system uniformly, reflecting general kinetic factors.
  \item \( \delta_{ij} \) is the Kronecker delta, ensuring that the metric is diagonal unless otherwise specified.
  \item \( f_{ij}(x) \) represents the interactions between species \( i \) and \( j \), dependent on their concentrations.
\end{itemize}

In this framework, the matrix \( g_{ij} \) describes how the concentrations of species interact within the reaction network. The value of \( \alpha \) is typically set based on the reaction rate constants or timescales of the network, while \( \beta \) adjusts the strength of the interaction between species. The functions \( f_{ij}(x) \) depend on the specific interactions, such as substrate-enzyme binding or product formation.

\subsection{Choosing the Values of \( \alpha \) and \( \beta \)}

\subsubsection*{1. \( \alpha \) – Scaling Factor}

Role: The constant \( \alpha \) determines the overall scaling of the system. It reflects kinetic factors such as the rate constants or timescales of reactions in the network.

Experimental Determination: \( \alpha \) can be determined through experimental reaction rates, such as in enzymatic reactions, where \( \alpha \) is proportional to the rate constant of the enzyme-substrate interaction.

\subsubsection*{2. \( \beta \) – Interaction Strength}

Role: \( \beta \) governs the strength of the interactions between species. It is a scaling factor for the interaction functions \( f_{ij}(x) \), which describe how species \( i \) and \( j \) influence each other in the network.

Experimental Determination: \( \beta \) can be determined by analyzing how the concentration of one species impacts another, and can be adjusted based on experimental data.

\subsection{Full Riemannian Metric Matrix Example}

For a system with four species, \( A \), \( B \), \( P \), and \( C \), where \( A \), and \( B \) are reactants; \( P \) is an intermediate; and \( C \) is the product.

The Riemannian metric matrix \( g_{ij}\) takes the form:

\[
g_{ij} = \begin{bmatrix}
g_{AA} & g_{AB} & g_{AP} & g_{AC} \\
g_{BA} & g_{BB} & g_{BP} & g_{BC} \\
g_{PA} & g_{PB} & g_{PP} & g_{PC} \\
g_{CA} & g_{CB} & g_{CP} & g_{CC}
\end{bmatrix}
\]

Where each element \( g_{ij} \) corresponds to the interaction between species \( i \) and \( j \). The diagonal terms represent self-interactions, and the off-diagonal terms represent cross-species interactions.

\subsection{Christoffel Symbols and Index \( l \) Calculation}

The Christoffel symbols represent how the curvature of the reaction network geometry dictates the way species concentrations evolve.

Formally, the Christoffel symbols \( \Gamma^i_{jk} \) are connection coefficients that describe the evolution of the reaction pathways in the curved space of the reaction network. These coefficients are derived from the metric tensor \( g_{ij} \), which encodes the geometry of the reaction space. These terms essentially quantify how each species' concentration impacts the overall network, similar to how curvature in space-time affects the motion of objects.

The general formula for the Christoffel symbols is given by:

\begin{equation}
\Gamma^i_{jk} = \frac{1}{2} g^{il} \left( \frac{\partial g_{lk}}{\partial x^j} + \frac{\partial g_{jl}}{\partial x^k} - \frac{\partial g_{jk}}{\partial x^l} \right)
\end{equation}

Here, \( i, j, k \) represent indices of species or dimensions in the configuration space, and the index \( l \) is the summation index that runs over all dimensions (or species) of the system.

This summation accounts for interactions between all species in the network and plays a crucial role in defining how the concentration of each species evolves with respect to others in the network.

\subsection{Understanding the Christoffel Symbol Formula}
An introduction into utilizing the differential geometry:

Metric Matrix \( g_{ij} \):
   The Christoffel symbols depend on the metric matrix \( g_{ij} \), which describes the geometry of the reaction space. In the case of a catalytic network with \( N \) species, this matrix could be of size \( N \times N \).

Inverse Metric Matrix \( g^{il} \):
   To compute the Christoffel symbols, we need the inverse of the metric matrix, denoted \( g^{il} \). The inverse metric is used to contract the Christoffel symbols. The inverse matrix \( g^{il} \) is crucial for the proper summation over the index \( l \).

Partial Derivatives:
   For each pair of indices \( (j, k) \), the formula requires computing partial derivatives of the metric components \( g_{jk} \), \( g_{jl} \), and \( g_{lk} \) with respect to the species or system parameters \( x^j \), \( x^k \), and \( x^l \).

Summing Over \( l \):
   The index \( l \) runs over all the possible dimensions or species in the reaction network. In a network with \( N \) species, \( l \) takes values \( 1, 2, \dots, N \). To calculate each Christoffel symbol, you must sum over the index \( l \), which involves summing over all possible dimensions of the system. This summation accounts for the interactions between all components of the network.

\subsection{Example: Computing a Christoffel Symbol}

For a system with three species \( A \), \( B \), and \( C \), the metric matrix \( g_{ij} \) might look like this:

\[
g_{ij} = \begin{bmatrix}
\alpha & \beta f_{AB}(x) & 0 \\
\beta f_{BA}(x) & \alpha & 0 \\
0 & 0 & \alpha
\end{bmatrix}
\]

Where:
\begin{itemize}
  \item \( \alpha \) is a constant that scales the system, reflecting general kinetic factors.
  \item \( f_{AB}(x) = x_A x_B \) and \( f_{BA}(x) \) represent the interactions between species \( A \) and \( B \), which depend on their concentrations.
  \item \( \beta \) adjusts the strength of the interaction between species.
\end{itemize}

To compute a Christoffel symbol such as \( \Gamma^1_{23} \), we perform the following steps:
\begin{itemize}
  \item Compute the inverse metric \( g^{il} \).
  \item Take the partial derivatives of \( g_{jk} \), \( g_{jl} \), and \( g_{lk} \) with respect to \( x^j \), \( x^k \), and \( x^l \).
  \item Sum over \( l \) to account for all dimensions.
\end{itemize}

For example, the Christoffel symbol \( \Gamma^1_{23} \) is calculated as:

\[
\Gamma^1_{23} = \frac{1}{2} g^{il} \left( \frac{\partial g_{l3}}{\partial x^2} + \frac{\partial g_{2l}}{\partial x^3} - \frac{\partial g_{23}}{\partial x^l} \right)
\]

The value of \( l \) runs over all possible species (here \( A \), \( B \), and \( C \)) and ensures that all interactions between the species are accounted for in the curvature of the reaction network.

\subsection{Key Points}

\begin{itemize}
  \item \( l \) is an index that runs over all the components (species or dimensions) of the system.
  \item The Christoffel symbols describe how the concentrations evolve in the curved space defined by the reaction network's geometry.
  \item Summing over the index \( l \) ensures that the interactions between all species in the network are considered when calculating the curvature and evolution of the system.
\end{itemize}

By computing the Christoffel symbols, we can understand how the chemical reaction pathways evolve within the constraints imposed by the curvature of the reaction space. This provides valuable insights into how the system self-organizes and saturates at equilibrium.

\subsection{Christoffel Symbols and Geodesic Equations}

Since Christoffel symbols \( \Gamma^i_{jk} \) capture how the reaction pathways evolve in the curved space defined by the metric, these symbols effectively modify the dynamics of the system, as the curvature of the configuration space affects how species concentrations evolve by directly impacting the expansion and constraint of reaction pathways. The geodesic equation describes this evolution:

\begin{equation}
\frac{d^2 x^i}{dt^2} + \Gamma^i_{jk} \frac{dx^j}{dt} \frac{dx^k}{dt} = F^i_{\text{chem}} + F_{\text{entropy}}(x)
\end{equation}

Where \( x^i \) represents the concentration of species \( i \), and \( F^i_{\text{chem}} \) is the chemical force term, describing the traditional reaction rates. The term \( F_{\text{entropy}}(x) \) is the entropy-like force that drives the system toward more probable configurations.

Constraints manifest as natural curvature bounds on geodesic expansion, enforcing structural limitations on reaction network evolution. From this, we find a key prediction of this model is that curvature constrains reaction trajectory expansion. This inherent constraint leads to saturation effects where reaction pathways self-limit beyond a critical complexity threshold. This naturally explains why real-world catalytic networks exhibit bounded diversity rather than unconstrained combinatorial growth.

\subsection{System-Specific Entropy-like Term in the Reaction Network}

Entropy, a thermodynamic quantity, measures the degree of disorder or probability of a system. In reaction networks, the system tends to evolve toward states of higher entropy. The system-specific entropy-like term \( F_{\text{entropy}}(x) \) is included to capture this behavior. It is defined as:

\begin{equation}
F_{\text{entropy}}(x) = \sum_i \frac{\partial S}{\partial x^i}
\end{equation}

Where \( S \) is a system-specific entropy-like function, which depends on the concentrations and interactions between species in the network. This term drives the system towards equilibrium by promoting configurations where entropy is maximized for the given network, ensuring that reaction pathways saturate when the system reaches its most probable state.

\subsection{Entropy and Self-Interruption}

In a reaction network, pathway saturation refers to the point at which the system reaches equilibrium, and no new viable reaction pathways can emerge. This can be viewed as a form of self-interruption, where the system reaches a steady-state or equilibrium. This saturation is directly influenced by the curvature of the reaction space. As the system evolves, the available pathways for further reactions narrow due to the curvature constraints imposed by the Riemannian metric.

The entropy-like force ensures that the reaction rates slow as the system approaches saturation, driving the system toward its final stable configuration. Eventually, the system reaches a stage where additional reactions no longer produce significant changes in the concentration of species. At this point, the network is self-limiting, and further expansion of reaction pathways becomes infeasible. This is the point of pathway saturation, where the system has exhausted all available reaction paths under the curvature constraints.

This behavior contrasts with traditional models where reaction spaces are unconstrained, and the system could continue expanding indefinitely. In our framework, the curvature-induced saturation effect constraints naturally enforce a bounded growth of the reaction network, leading to a saturation point where further reaction pathways cease to significantly affect the system’s state.

\subsection{Geodesic Deviation and Stability of Reaction Networks}
The stability of competing reaction pathways can be analyzed through the geodesic deviation equation:
\begin{equation}
    \frac{D^2 \xi^i}{Dt^2} + R^i_{jkl} \frac{dx^j}{dt} \xi^k \frac{dx^l}{dt} = 0,
\end{equation}
where \( R^i_{jkl} \) is the Riemann curvature tensor induced by reaction constraints. If \( R^i_{jkl} > 0 \) in certain regions of \( \mathcal{C} \), chemical pathways experience divergence and bifurcation, leading to distinct network topologies and emergent phase transitions in reaction complexity, i.e., the system splits into different network topologies as the complexity of the network grows. \(D\) is capitalized, somewhat arbitrarily, per the approach offered by Misner, Wheeler, and Thorne (1973).

\section{Simulation of the Geometrical Approach to Reaction Network Analysis}

The proposed model can be numerically simulated by solving the system of ordinary differential equations (ODEs) that describe the evolution of species concentrations. The system of ODEs is given by \(Eq.3\), where the terms \( \Gamma^i_{jk} \), \( F^i_{\text{chem}} \), and \( F_{\text{entropy}}(x) \) are calculated as described earlier. The system can be solved numerically using methods such as Runge-Kutta or Euler’s method.

The system of ODEs given by \(E1.3\)  describes the evolution of species concentrations over time, and the numerical methods chosen (such as Runge-Kutta or Euler’s method) are necessary to simulate the complex, nonlinear interactions dictated by the Riemannian metric.

\subsection{Example Simulation Setup}

For the system with three species, we set the Riemannian metric \( g_{ij} \) as:

\[
g_{ij} = \begin{bmatrix}
\alpha & \beta f_{AB}(x) & 0 \\
\beta f_{BA}(x) & \alpha & 0 \\
0 & 0 & \alpha
\end{bmatrix}
\]

Where:
\begin{itemize}
  \item \( \alpha \) is a constant that scales the system, reflecting general kinetic factors.
  \item \( f_{AB}(x) = x_A x_B \) and \( f_{BA}(x) \) similarly represent the interactions between species \( A \) and \( B \), which depend on their concentrations.
  \item \( \beta \) adjusts the strength of the interaction between species.
\end{itemize}
 The Riemannian metric is defined by interaction terms such as:

\[
g_{AB} = \beta f_{AB}(x) = \beta x_A x_B
\]

 The system is solved over time, with concentration profiles being plotted for each species.

The evolution of the system is governed by the geodesic equation, which describes how the concentration of species evolves under the curvature of the reaction space given once more by \(Eq.3\):

\[
\frac{d^2 x^i}{dt^2} + \Gamma^i_{jk} \frac{dx^j}{dt} \frac{dx^k}{dt} = F^i_{\text{chem}} + F_{\text{entropy}}(x)
\]

As the system evolves, the concentrations of \( A \), \( B \), and \( C \) change according to the geodesic equation. However, due to the curvature of the reaction space, the system eventually reaches a point where no new reaction pathways can emerge. The available pathways become saturated, and the system approaches equilibrium.

Before saturation: The concentrations evolve freely, but as the system progresses, the available "pathways" for further reactions start to narrow due to the curvature constraints imposed by the Riemannian metric. The rate of change in concentrations slows as the system approaches saturation.

At saturation: The reaction pathway becomes limited, and additional reactions no longer produce significant changes. This is the point of pathway saturation, where the system has exhausted all available reaction paths under the constraints imposed by curvature.

The entropy-like term, \( F_{\text{entropy}}(x) \), also plays a key role in driving the system toward saturation. As the concentrations of the species evolve, the system tends to shift toward higher-entropy states, which correspond to configurations where the concentration of \( C \) is maximized.

This term is given by \(Eq.4\) , where \( S \) is a system-specific entropy function, which depends on the concentrations and interactions between species in the network.

The key prediction of this model is that our analyzed curvature constrains the growth of reaction pathways and leads to pathway saturation at some certain point. As the reaction network evolves, the increasing curvature in the configuration space leads to diminishing opportunities for new reaction pathways. This constraint eventually forces the system to reach a saturation point, at which no further significant reactions can occur, thereby stabilizing the network. This mode of analysis is fundamentally distinct from traditional models that treat reaction space as unconstrained, where the system could continue expanding indefinitely. 

\subsection{Application to Catalytic Network Saturation}
To demonstrate additional applicability of this method, we set up a catalytic system where reaction intermediates exhibit recursive production-depletion cycles. The effective pathway curvature is determined by:
\begin{equation}
    R = \frac{1}{2} \left( \frac{\partial^2 S}{\partial x^i \partial x^j} - g^{ij} \Gamma^k_{ij} \frac{\partial S}{\partial x^k} \right),
\end{equation}
where \( S \) represents an entropy-like constraint function measuring network redundancy. 

For example, in enzymatic catalysis, saturation occurs when all active sites are occupied, limiting further reaction progress. Our model provides a mathematical description of this behavior in terms of network curvature.

This formulation allows for: Prediction of saturation thresholds—where additional reaction pathways fail to contribute new chemical diversity; identification of structurally constrained reaction motifs, explaining why some catalytic cycles dominate over others; and analysis of robustness—showing how perturbations induce structural shifts in the network.

While the above is presented in brief, it aims to show the possibility for additional applications beyond the introductory scope of the overarching framework presented in this paper.

\section{Discussion}

\subsection{Scalability and Computational Considerations}

This framework is designed to be scalable and can be applied to larger, more complex reaction networks, such as those found in metabolic systems or catalytic cycles. However, as the number of species and reactions increases, the system of equations becomes more computationally challenging to solve due to the increased dimensionality and complexity of the Riemannian metric and Christoffel symbols.

Computational methods such as parallel computing, sparse matrix techniques, and graph-based algorithms can be used to handle these larger systems. Despite the computational challenges, this framework provides a scalable approach for modeling systems of increasing complexity.

\subsection{Unique and Beneficial Aspects}

Unlike conventional methods which treat reaction networks as combinatorially unconstrained, we consider how network topology and species interactions emerge as structural limits, inherently constrained by the curvature of the system's manifold. This enables a more accurate modeling of the emergent dynamics of pathway saturation.

While previous work, such as the framework proposed by Hirono et al. (2021), focuses on structural reduction of reaction networks based on network topology and stoichiometry--or Hessian analysis (Kobayashi 2022)--providing a more computationally efficient approach for large systems, our framework provides a deeper exploration of network complexity. By introducing differential geometry, we model the curvature-induced constraints that limit reaction pathways and dictate saturation. This gives us a more detailed understanding of how complex systems evolve over time, offering insights that traditional models may overlook.

Our approach provides a unique perspective on reaction networks by combining traditional reaction kinetics with the geometrical constraints of differential geometry. The introduction of the entropy-like term and curvature effects allows the model to capture emergent behaviors such as pathway saturation, which are not readily observable in traditional kinetic models. Additionally, the framework's ability to model self-interruption and thermodynamic equilibrium is particularly useful in understanding the long-term dynamics of complex systems.

\section{Conclusion: A Localized Framework for Chemical Complexity Analysis}
By analyzing curvature constraints of the differential geometry of reaction networks, we obtain a predictive tool for analyzing chemical reaction complexity growth without requiring exhaustive combinatorial searches. The geometric framework presented here allows for a structured understanding of reaction pathway evolution and saturation effects.

This approach provides an alternative to purely kinetic models, offering new insights into how reaction networks self-organize under intrinsic constraints. By introducing curvature constraints through integrating the Riemannian metric, Christoffel symbols, and entropy-like terms, we can capture both the chemical kinetics and the geometrical constraints that govern the system's dynamics. This model is scalable and applicable to a wide range of systems, from biochemical networks to industrial catalysis, providing a richer understanding of how these networks evolve over time. Future work will aim to integrate Lyapunov stability analysis and Perturbation theory approaches.

\end{document}